\RequirePackage[l2tabu,orthodox]{nag}
\documentclass[letterpaper,peerreview,12pt,draftclsnofoot]{IEEEtran}
\usepackage{amsmath,amsfonts,amssymb}
\usepackage{array}
\usepackage[caption=false,font=normalsize,labelfont=sf,textfont=sf]{subfig}
\usepackage{textcomp}
\usepackage{stfloats}
\usepackage{url}
\usepackage{verbatim}
\usepackage{graphicx}
\usepackage{cite}
\usepackage[inline]{enumitem}
\usepackage[table,svgnames]{xcolor}
\usepackage{algorithm}
\usepackage{hyperref}
\usepackage{orcidlink}

\newtheorem{lem}{Lemma}
\newtheorem{thm}{Theorem}

\newtheorem{prop}{Proposition}

\newtheorem{ex}{Example}

\newcommand{\vek}[1]{\boldsymbol{\mathbf{#1}}}
\makeatletter
\newcommand{\bra}[1]{\langle#1\@ifnextchar\ket{}{|}}
\makeatother
\newcommand{\ket}[1]{|#1\rangle}
\newcommand{\symp}[2]{(#1|#2)}
\def\F{\mathbb{F}}
\def\tr{\mathop{\mathrm{tr}}}

\begin{document}

\title{Puncturing Quantum Stabilizer Codes}

\author{Jaron Skovsted Gundersen${}^{\orcidlink{0000-0003-0882-4621}}$,
    René Bødker Christensen${}^{\orcidlink{0000-0002-9209-3739}}$, \\
      Markus Grassl${}^{\orcidlink{0000-0002-3720-5195}}$, \IEEEmembership{Senior Member, IEEE},
      Petar Popovski${}^{\orcidlink{0000-0001-6195-4797}}$,~\IEEEmembership{Fellow, IEEE}, and\\ Rafał Wisniewski${}^{\orcidlink{0000-0001-6719-8427}}$,~\IEEEmembership{Senior Member, IEEE}.
    \thanks{J.S. Gundersen (jaron@es.aau.dk),  P. Popovski (petarp@es.aau.dk), and R. Wisniewski (raf@es.aau.dk) are with the Department of Electronic Systems at Aalborg University.}
    \thanks{R.B. Christensen (rene@math.aau.dk) is with the Department of Mathematical Sciences and the Department of Electronic Systems, both at Aalborg University.}
    \thanks{M. Grassl (markus.grassl@ug.edu.pl) is with the
      International Centre for Theory of Quantum Technologies at the University of Gdansk.}%
    }

\markboth{Puncturing Quantum Stabilizer Codes}{Puncturing Quantum Stabilizer Codes}


\maketitle

\begin{abstract}
  Classical coding theory contains several techniques to obtain new
  codes from other codes, including puncturing and shortening. Both of these techniques have been generalized to quantum codes. Restricting to stabilizer codes, this paper introduces
  more freedom in the choice of the encoded states after puncturing. Furthermore, we also give an explicit description of the stabilizers for the punctured code.
  The additional freedom in the procedure also opens up for new ways to
  construct new codes from old, and we present several ways to utilize
  this in the search for codes with good or even optimal parameters.
  In particular, we use the construction to obtain codes whose parameters
  exceed the best previously known and which are better than what general puncturing guarantees.
  Lastly, the freedom in our puncture procedure allowed us to generalize the proof of the Griesmer bound from the classical
  setting to stabilizer codes for qudits of prime dimension since the proof relies heavily on the
  puncturing technique.
\end{abstract}


\section{Introduction}
\IEEEPARstart{T}{wo} important and recurring themes within coding theory are
\begin{enumerate*}[label=(\alph*),itemjoin*={{, and }}]
  \item \label{item:codes} the construction of error-correcting codes with good parameters
  \item \label{item:bound} determining the optimal parameters of error-correcting codes.
\end{enumerate*} 
In that sense minimizing the gap between achievable parameters from
known constructions and theoretical upper bounds on their parameters
is a major research topic. See for instance \cite{Grassl:codetables}
for an overview of the strictest bounds and the best known parameters.

While \ref{item:bound} is achieved by proving bounds on the possible
parameters, \ref{item:codes} is typically done by using some algebraic
or combinatorial structure to construct a good code. Another approach
is to start with a good code and then to modify it by using various
techniques. Two of the most well-known techniques of this kind are
\emph{shortening} and \emph{puncturing}.  The hope is that these new
codes have good parameters since they inherit some of the structure
and properties of the initial code.

For classical codes, puncturing an $[n,k,d]_q$ code $C$ removes an
entry from every codeword. As long as the minimum distance $d$ of $C$
is at least $2$, the punctured code will have parameters
$[n-1,k,d']_q$, where $d'=d$ or $d'=d-1$, depending on the choice of
punctured entry and the structure of the code.  For quantum stabilizer
codes, it is also possible to perform a type of
puncturing \cite{Grassl21}, but an extra step is required. Rather than
simply removing a qudit from every encoded state, we first force that
qudit to be in a specific state before removing it. In this paper, we
show that puncturing a stabilizer code can be seen as a generalization
of a shortening in the classical setting.

Puncturing of general quantum codes is presented in \cite{Grassl21},
focusing on the code space. In this work, we restrict ourselves to
stabilizer codes, which allows us to take a different approach and
describe puncturing from the viewpoint of the stabilizer
representation using symplectic vectors. We show how this allows
puncturing to be performed directly on the stabilizer matrix, thus
explicitly obtaining the generators of the punctured stabilizer. A
similar strategy can be found in \cite[Theorem 6 d)]{CodesF4} for qubits, without
addressing the variety of choices. Here, we consider the
general case of qudits and discuss how the puncturing procedure actually
allows additional freedom that has not been previously described in the literature.

Our insight into the puncturing procedure also provides novel ways to
design new codes with good or even optimal parameters, starting
from the best currently known codes. We present these ideas in
Sections~\ref{sec:Idea_to_good_codes} and \ref{sec:shortening}.
Furthermore, we apply these techniques in Section~\ref{Sec:New_codes}
to produce new codes whose parameters exceed the best known codes listed in~\cite{Grassl:codetables}.

Finally, we show that our description of puncturing may sometimes be
used to transfer classical proofs to the quantum setting. More
precisely, we show that the proof of the well-known Griesmer
bound \cite{Griesmer60,huffman2003fundamentals} for classical codes
can be generalized to obtain a quantum Griesmer bound.  Several
Griesmer type bounds already exist for quantum error-correcting codes.
Sarvepalli et al.~\cite{Sarvepalli10} proved a Griesmer bound for CSS
codes, and Luo et al.~\cite{Luo22} have extended this bound to the
asymmetric CSS construction.  Our bound is comparatively weak, but
holds for all stabilizer codes over $\F_p$, where $p$ is a prime.

In summary, the contributions of the current work may be described succinctly as
\begin{itemize}
  \item A generalized puncturing of quantum stabilizer codes that allows
  more flexibility compared to previous procedures.
  We give an explicit description of the procedure based on the stabilizer
  matrices, directly describing the stabilizers after puncturing;
  \item A description of ways in which this puncturing technique can be used to
  construct new quantum error-correcting codes whose parameters improve
  the best known;
  \item Translation of the classical proof of the Griesmer bound to
    its quantum analog. 
\end{itemize}

From a system-level perspective, an information system -- be it for communication or computation -- may be required to support codes of different rates in order to operate efficiently in different error regimes.
The system design could in principle specify each such code explicitly. With puncturing, however, a large family of code rates may be obtained by specifying a small number of `mother' codes paired with multiple descriptions of puncturing operations.
The increased flexibility of our proposed method allows a wider range of code parameters in this type of setting.

Before presenting our main results, we would like to note that for quantum codes, the existence of a code with parameters $[\![n,k,d]\!]_q$ does not necessarily guarantee the existence of a code with parameters $[\![n-1,k-1,d]\!]_q$. In terms of parameters, that would be the analog of shortening.
The so-called \emph{puncture code} (see \cite{rains}) characterizes when shortening a quantum code of length $n$ to one of length $n-s$ is possible. Despite the name \emph{puncture code}, this should not be confused with puncturing of a quantum code.

This work is structured as follows. Section~\ref{sec:preliminaries}
summarizes the notation used for classical codes and the theory of
stabilizer codes that will be used in later sections.
Section~\ref{sec:punct-stab} addresses puncturing of stabilizer codes
in the symplectic vector representation and its translation to
stabilizer matrices. It also describes possible strategies for
searching for good codes, and relates our puncturing description to
those that already exist in the literature. Afterwards,
Section~\ref{sec:quant-griesm-bound} applies the puncturing procedure
to obtain a quantum analog of the Griesmer bound. In
Section~\ref{sec:random-punct}, we perform numerical experiments to
gauge how often successively punctured codes have minimum distances
that exceed the worst-case bound.
In Section~\ref{Sec:New_codes}, we apply the techniques of
Section~\ref{sec:punct-stab} to give examples of new quantum
error-correcting codes that have better parameters than previously
known constructions. Section~\ref{sec:concl-open-probl}
contains the conclusion.

\section{Preliminaries}\label{sec:preliminaries}
Throughout this work, let $p$ be a prime number, and let
$\F_q$ be the finite field with $q=p^r$ elements.  We present
classical codes for general $q$ since extension fields do not
complicate the presentation.  When describing quantum error-correcting
codes, however, we will mainly restrict ourselves to the prime field
case, i.\,e., $q=p$. This simplifies certain parts of the description,
and for the puncturing in Section~\ref{sec:punct-stab} we mainly discuss
the prime field case.  We focus on this case because it allows us to
explicitly describe those elements that are carried over to the
punctured code.

\subsection{Classical linear codes}
We briefly recall relevant concepts from classical coding theory. A
detailed treatment can be found in textbooks such
as~\cite{MacWilliamsSloane,huffman2003fundamentals}.

A linear error-correcting code is a $k$-dimensional subspace
$C\subseteq \F_q^n$. It can be represented by a $k\times n$
generator matrix $G$, which maps $\mathbf{m}\in\F_q^k$ to $C$
via the matrix product $\mathbf{m}G$. For any element $\mathbf{x}\in\F_q^n$, its Hamming
weight $w_H(\mathbf{x})$ is defined as the number of nonzero
entries. Based on this, the Hamming distance between two elements
$\mathbf{x},\mathbf{y}\in\F_q^n$ is
\begin{equation*}
    d_H(\mathbf{x},\mathbf{y})=w_H(\mathbf{x}-\mathbf{y}) = |\{ i\mid x_i\neq y_i\}|.
\end{equation*}
For error-correction purposes, the minimum distance $d=\min_{\mathbf{x},\mathbf{y}\in C,\mathbf{x}\neq\mathbf{y}} d_H(\mathbf{x},\mathbf{y})$ is of particular interest. The
 linearity of $C$ implies that $d$ is in fact the minimum weight of the nonzero codewords in $C$.
With these parameters, we call $C$ an $[n,k,d]_q$ code.

Certain operations modify linear codes in an attempt to obtain new
codes with good parameters while preserving linearity. Two well-known
operations of this kind are \textit{puncturing} and
\textit{shortening}. For puncturing, we introduce the map
\begin{align}\label{eq:puncture}
    \pi_I(\vek{c})=(c_j)_{j\notin I},
\end{align}
which removes the entries in the vector $\vek{c}$ corresponding to the
indices in the set $I$. When $I=\{i\}$, we will often abuse the notation and write $\pi_i$.

Extending this idea to a set $C$, we
use the notation $\pi_I(C)=\{\pi_I(\vek{c})\mid \vek{c}\in C\}$. If
$C$ is a linear code, $\pi_I(C)$ is also linear and is called a
punctured code. Furthermore, if $|I|<d$ and $C$ has parameters
$[n,k,d]_q$, then the punctured code is an $[n-|I|,k,d']_q$-code,
where $d'\geq d-|I|$.

The shortening of a code $C$ at an index $i$ comprises a restriction
to those codewords $\mathbf{c}\in C$ with $c_i=0$ followed by a
puncturing on the $i$-th index. If every codeword has $c_i=0$, then
shortening and puncturing will have the same effect on the code, and
the resulting parameters are $[n-1,k,d]_q$. If, on the other hand,
$d>1$ and some codeword has $c_i\neq 0$, then shortening the $i$-th
coordinate produces a code of parameters $[n-1,k-1,d]_q$.

To simplify notation, in what follows we will only consider puncturing
and shortening in the first entry.  In preparation for the quantum
case, shortening can also be described by considering the map
\begin{align}\label{eq:shortening_map}
  \sigma_a\colon\left\{
  \begin{array}{@{}r@{}c@{}l}
    C & {}\rightarrow{} & \F_q\\
    \mathbf{c}&\mapsto&ac_1
  \end{array}
  \right.
\end{align}
where $a\in \F_q$. The map projects onto the first coordinate of each codeword and multiplies the result by $a\in\mathbb{F}_q$. For $a\neq 0$, shortening of $C$
produces the code $\pi_1(\ker\sigma_a)$. That is, shortening is a
restriction to $\ker\sigma_a$ followed by a puncturing.

\subsection{Stabilizer codes}\label{sec:stabilizer-codes}
    We
    briefly introduce the theory needed for describing stabilizer
    codes. For a more in-depth introduction, see
    e.\,g., \cite{Nonbinary_Stabilizer}.

A single
qudit is described by an element of $\mathbb{C}^q$, and we fix some
orthonormal basis $\{|i\rangle\mid i \in \F_q\}$ of
$\mathbb{C}^q$ labelled by the elements of $\F_q$. We sometimes switch to Greek letters to avoid interpretation of Roman letters as integers. To describe the errors affecting a qudit, define the 
(generalized) bit flip and phase shift errors represented by the
operators
\begin{align*}
    X(a)&\colon |i\rangle \mapsto |i+a\rangle\qquad\text{and}\\
    Z(b)&\colon |i\rangle \mapsto \omega^{\tr(ib)}|i\rangle,
\end{align*}
respectively, where $\omega\in \mathbb{C}$ is a primitive $p$-th root
of unity and $\tr$ is the (absolute) trace from $\F_q$
to $\F_p$ given by $\tr(a) = \sum^{r-1}_{i=0} a^{q^i}$.
When $q=p$ is prime, the trace is the identity on
$\F_p$ and can be omitted.


    The bit flip errors may be extended to $n$-qudit errors by defining $X(\vek{a})=\bigotimes_{i=1}^n X(a_i)$, and the $n$-qudit phase shift $Z(\mathbf{b})$ may be defined similarly.
    One may then show, see e.g. \cite{Nonbinary_Stabilizer}, that all $n$-qudit errors can be described by the multiplicative group
\begin{align}\label{eq:errorGroup}
  \mathcal{G}_n&=\{ \omega^cX(\vek{a})Z(\vek{b})\mid \vek{a},\vek{b}\in \F_q^n, c\in \F_p\} .
\end{align}
For even characteristic (i.e. $p=2$), the authors of~\cite{Ball_2023} suggest that
    \begin{align*}
      \mathcal{G}_n&=\{ \omega^ci^bX(\vek{a})Z(\vek{b})\mid \vek{a},\vek{b}\in \F_{2^r}^n,\; b,c\in \F_2\} \\
                   &=\bigr\{ i^m X(\vek{a})Z(\vek{b})\mid \vek{a},\vek{b}\in \F_{2^r}^n, m\in\{0,1,2,3\}\bigl\} 
    \end{align*}
    is a more natural choice.
    This is based on the observation \cite[Lem.~5.2]{Ball_2023} that the possible eigenvalues of $X(a)Z(b)$ are  the $p$-th roots of unity $\{\omega^c\mid c\in\mathbb{F}_p\}$ for odd characteristic, but $\{\omega^ci^b\mid c,b\in\mathbb{F}_2\}$ for even characteristic.

$\mathcal{G}_n$ is a finite group of order $p\cdot q^{2n}$
($4\cdot q^{2n}$ for $q=2^r$) even though infinitely many
errors exist. However, due to the way measurements are used in the
correction procedure, it is enough to focus on $\mathcal{G}_n$, see
for instance \cite{gottesman1997stabilizer,KnillLaflamme}.

Although $\mathcal{G}_n$ contains a global phase factor, it will not
have an effect on measurement results during
error-correction. Therefore, it is common to focus not on
$\mathcal{G}_n$, but on the quotient $\mathcal{G}_n/\langle\omega
I\rangle$ ($\mathcal{G}_n/\langle iI\rangle$ for even
characteristic). This quotient is isomorphic to the additive group
$\F_q^{2n}$ via the isomorphism $\varphi\colon
\mathcal{G}_n/\langle\omega I\rangle \to \F_q^{2n}$ given by
\begin{align}\label{eq:iso_to_field}
  \varphi\colon\left\{
  \begin{array}{@{}r@{}c@{}l}
    \mathcal{G}_n/\langle\omega I\rangle & {}\rightarrow{} &\F_q^{2n}\\
    \omega^cX(\mathbf{a})Z(\mathbf{b}) & \mapsto & \symp{\mathbf{a}}{\mathbf{b}}
  \end{array}
  \right.
\end{align}
for $q=p^r$, $p>2$. The case $q=2^r$ is similar, but with $i^m$
replacing $\omega^c$.

The abstraction from quantum errors to $\F_q^{2n}$ is for
instance useful in gauging the severity of an error, i.e. the number of qudits affected.
With the map from~\eqref{eq:iso_to_field}, 
elements $\omega^cX(\mathbf{a})Z(\mathbf{b})$ in the same equivalence class
in $\mathcal{G}_n/\langle \omega I\rangle$ have the same symplectic
weight defined as
\begin{align*}
    w_s(\vek{a}|\vek{b})=|\{i\mid a_i\neq 0 \vee b_i\neq 0\}|.
\end{align*}
If not explicitly stated otherwise, in the context of stabilizer codes
the term `weight' will always refer to this symplectic weight.

    An $[\![n,k]\!]_q$ quantum error-correcting code of length $n$ is a
    $q^k$-dimensional subspace of $\mathbb{C}^{q^n}\cong \mathbb{C}^q\otimes\dots\otimes\mathbb{C}^q$. One common way to obtain such quantum error-correcting codes is via the
stabilizer formalism.  A stabilizer code is a subspace $\mathcal{C}_S\subseteq
\mathbb{C}^{q^n}$ associated with an Abelian subgroup $S\subseteq
\mathcal{G}_n$ that intersects trivially with the center of
$\mathcal{G}_n$.  The stabilizer code is the joint $+1$-eigenspace of
the elements in $S$, i.\,e.,
\begin{equation}\label{eq:stabilizerDef}
  |\psi\rangle \in \mathcal{C}_S
  \quad\Longleftrightarrow\quad
  \forall M\in S\colon M|\psi\rangle=|\psi\rangle.
\end{equation}
For $q=p^r$, if $S$ has $r(n-k)$ independent generators, the
stabilizer code has dimension $q^n/q^{n-k}=q^k$, and thus we obtain an
$[\![n,k]\!]_q$ stabilizer code.  We encode $k$ qudits into a linear
combination of $q^k$ mutually orthogonal states in $\mathcal{C}_S$. Hence, if no
errors occur we will stay in $\mathcal{C}_S$.  Contrary to the classical setting,
however, some errors will not affect the encoded states. These
are exactly the errors corresponding to the elements in $S$ since for
all $M\in S$ and $|\psi\rangle \in \mathcal{C}_S$, we have
$M|\psi\rangle=|\psi\rangle$.

Thus, errors from $S$ have no effect, and are hence `corrected'
automatically in a stabilizer code. The full set of correctable errors may be determined from the Knill-Laflamme conditions~\cite{KnillLaflamme}, which imply that the minimum distance $d$ of a stabilizer code $\mathcal{C}_S$ is
equal to the minimum weight of the non-scalar elements in
$C(S)\setminus S$.  Here, $C(S)$ denotes the centralizer of $S$ in $\mathcal{G}_n$. When $C(S)$ contains an element of weight strictly
smaller than $d$, that element is also in $S$, and the code is called
impure; otherwise it is called pure. For additive codes, note that the terms pure and impure correspond exactly to the terms non-degenerate and degenerate, respectively~\cite{CodesF4}.
When $C(S)=\langle \omega I,S\rangle$, the
dimension of $\mathcal{C}_S$ is one, and the minimum distance is given by the
minimum weight of non-scalar elements in $S$.

We will use the
notations $S_q^\perp$ and $S_q$ for the subspaces of $\F_q^{2n}$
obtained when applying the isomorphism in \eqref{eq:iso_to_field} on
$C(S)$ and $S$, respectively. That is, $S_q=\varphi(S)$ and
$S_q^\perp=\varphi(C(S))$. The reason for this notation is that
$S_q^\perp$ is exactly the dual of $S_q$ with respect to the trace-symplectic form
\begin{equation}\label{eq:symplectic}
    \langle(\vek{a}|\vek{b}),(\vek{a}'|\vek{b}')\rangle_s=\tr\left(\langle \vek{b}, \vek{a}'\rangle-\langle\vek{b}', \vek{a}\rangle\right),
\end{equation}
where $\langle \cdot,\cdot \rangle$ is the standard Euclidean inner
product. 
Elements from $\mathcal{G}_n$ form a minimal generating set
for $S$ if and only if the corresponding elements in $\F_q^{2n}$ are
additively independent. Note that when $S_q$ is an
$\F_q$-linear vector space, then the trace in
\eqref{eq:symplectic} can be omitted.  A group $S$ is Abelian if and
only if the trace symplectic product of any two elements in $S_q$ is
zero. Let $S=\langle M_1,\ldots,M_{r(n-k)}\rangle$, and assume that the generators $M_i$ are independent. The vectors $\{\varphi(M_i)\}$
are then additively independent and constitute a basis for the
additive code $S_q$. Let $G_S$ be the $r(n-k)\times 2n$ matrix over $\mathbb{F}_q$ having $\varphi(M_i)$ as
$i$-th row. We refer to $G_S$ as a stabilizer matrix.  Since $S$ is
Abelian, we have $S_q\subseteq S_q^\perp$. Thus, by adding an additional $2rk$
rows to $G_S$, denoted by $\widetilde{G}_{S^\perp}$, we obtain a matrix whose rows generate
$S_q^\perp$. We call this the centralizer matrix, and we will
use the notation
\begin{equation*}
  G=\left[\begin{array}{c}
    G_S\\
    \hline
    \widetilde{G}_{S^\perp}
  \end{array}\right].
\end{equation*}

\section{Puncturing a stabilizer}\label{sec:punct-stab}
A general puncturing procedure has already been described
in~\cite[Sec.~4]{Grassl21} from the point of view of the stabilized code space
$\mathcal{C}_S$.  We take a slightly different approach and focus on the
symplectic vectors in $\F_q^{2n}$, and show that this allows
us to directly compute the centralizer matrix for the
punctured code. In Section~\ref{sec:relationGrassl}, we then relate
this type of puncturing to the one from \cite{Grassl21}.

\subsection{A symplectic viewpoint}\label{sec:punctureSymplectic}
Let $S$ be a stabilizer, and let $G$ be its centralizer matrix
as described in Section~\ref{sec:stabilizer-codes}.  To simplify
notation, we focus on the case $q=p$ prime; the case $q=p^r$, $r>1$
will be addressed in Section \ref{sec:nonprime}.  We will consider
elements $\symp{\mathbf{a}}{\mathbf{b}}$ in the space
$\F_p^{2n}$ as vectors of length $n$ having pairs as entries
$\symp{a_i}{b_i}$.  Therefore, when we later apply $\pi_I$ as
introduced in the classical setting in \eqref{eq:puncture} to
$\symp{\vek{a}}{\vek{b}}\in \F_p^{2n}$, we are removing
$(a_i|b_i)$ for $i\in I\subseteq \{1,2,\ldots,n\}$ and obtain an
element in $\F_p^{2(n-|I|)}$.  In the following, we will always describe
puncturing on the first index, which corresponds to columns $1$ and
$n+1$ in $G$. This eases notation, but we note that this procedure can
be applied to any index.

To perform puncturing, choose some nonzero
$\symp{\alpha}{\beta}\in\F_p^2\setminus \{\symp{0}{0}\}$ and
define the punctured stabilizer
$S_p^{\symp{\alpha}{\beta}}\subseteq\F_p^{2(n-1)}$ as
\begin{equation}\label{eq:puncturedStabilizer}
  S_p^{\symp{\alpha}{\beta}} = \{ \pi_1\symp{\mathbf{a}}{\mathbf{b}} \mid \symp{\mathbf{a}}{\mathbf{b}}\in S_p, \langle \symp{a_1}{b_1},\symp{\alpha}{\beta}\rangle_s=0 \}.
\end{equation}
In other words, $S_p^{\symp{\alpha}{\beta}}$ contains the projections
of elements in $S_p$ that are symplectically orthogonal to
$\symp{\alpha}{\beta}$ on the first index. We will use the map
$\sigma_{\symp{\alpha}{\beta}}\colon S_p\rightarrow\F_p$ given
by
\begin{align}\label{eq:symplectic_sigma}
  \sigma_{\symp{\alpha}{\beta}}\symp{\vek{a}}{\vek{b}} =
    \langle \symp{a_1}{b_1},\symp{\alpha}{\beta}\rangle_s,
\end{align}
since $S_p^{\symp{\alpha}{\beta}}$ can then be described as
$\pi_1(\ker\sigma_{\symp{\alpha}{\beta}})$. Note here the similarities
to shortening of a classical code, where we introduced the map
$\sigma_a$ in \eqref{eq:shortening_map}. Therefore, a puncturing of a
quantum stabilizer code can be carried out more or less by making a
classical shortening on the stabilizer matrix. And as we show below in
Proposition~\ref{prop:dualSpP}, we can also carry out the same
kind of shortening on the centralizer matrix to obtain the
centralizer for the punctured code.
  
  Since we are considering vectors over $\mathbb{F}_p$ we immediately obtain the following lemma.
\begin{lem}\label{lem:puncture}
  For
    $\symp{\alpha}{\beta}\ne\symp{0}{0}$, the condition $\langle
    \symp{a_1}{b_1},\symp{\alpha}{\beta}\rangle_s=0$ is equivalent to
    $\symp{a_1}{b_1}=c\symp{\alpha}{\beta}$ for some
    $c\in\F_p$.
\end{lem}
We will use this fact several times in this work. This viewpoint makes
it easy to prove that $S_p^{\symp{\alpha}{\beta}}$ does in fact
represent a stabilizer itself.
\begin{prop}
  Let $S_p\subseteq \F_p^{2n}$ represent a stabilizer. Then
  $S_p^{\symp{\alpha}{\beta}}$ is self-orthogonal with respect to the
  symplectic product.
\end{prop}
\begin{IEEEproof}
  Let $\symp{\mathbf{u}}{\mathbf{v}}$ and
  $\symp{\mathbf{u}'}{\mathbf{v}'}$ be two elements of
  $S_p^{\symp{\alpha}{\beta}}$. By definition, there must be some
  $a,b\in\F_p$ such that
  $\pi_1\symp{a,\mathbf{u}}{b,\mathbf{v}}=\symp{\mathbf{u}}{\mathbf{v}}$
  and $\langle\symp{a}{b},\symp{\alpha}{\beta}\rangle_s=0$. Similarly,
  $a',b'\in\F_p$ can be found for
  $\symp{\mathbf{u}'}{\mathbf{v}'}$.
  Using Lemma \ref{lem:puncture}, $\langle\symp{a}{b},\symp{\alpha}{\beta}\rangle_s=\langle\symp{a'}{b'},\symp{\alpha}{\beta}\rangle_s=0$ implies that $\symp{a}{b}=c\symp{\alpha}{\beta}$ and $\symp{a'}{b'}=c'\symp{\alpha}{\beta}$, and hence $\langle\symp{a}{b},\symp{a'}{b'}\rangle_s=0$
  Then
  \begin{equation*}
    \langle \symp{\mathbf{u}}{\mathbf{v}}, \symp{\mathbf{u}'}{\mathbf{v}'}\rangle_s
    = \langle\symp{a}{b},\symp{a'}{b'}\rangle_s+\langle \symp{\mathbf{u}}{\mathbf{v}}, \symp{\mathbf{u}'}{\mathbf{v}'}\rangle_s
    = \langle\symp{a,\mathbf{u}}{b,\mathbf{v}}, \symp{a',\mathbf{u}'}{b',\mathbf{v}'}\rangle_s = 0,
  \end{equation*}
  where the last equality comes from $S_p$ being a stabilizer.
\end{IEEEproof}

\begin{prop} \label{prop:dualSpP}
  Let $S_p^{\symp{\alpha}{\beta}}$ be as in~\eqref{eq:puncturedStabilizer}. Then
  \begin{equation*}
    (S_p^{\symp{\alpha}{\beta}})^\perp = (S_p^\perp)^{\symp{\alpha}{\beta}}
  \end{equation*}
\end{prop}
\begin{IEEEproof}
  First, let $\symp{\mathbf{u}}{\mathbf{v}}\in (S_p^\perp)^{\symp{\alpha}{\beta}}$, meaning that
  $\symp{a,\mathbf{u}}{b,\mathbf{v}}\in S_p^\perp$ for some $a,b\in\F_p$ with $\langle\symp{a}{b},\symp{\alpha}{\beta}\rangle_s=0$.
  We then have $\langle \symp{a,\mathbf{u}}{b,\mathbf{v}},
  \symp{a',\mathbf{u}'}{b',\mathbf{v}'}\rangle_s=0$ for every
  $\symp{a',\mathbf{u}'}{b',\mathbf{v}'}\in S_p$. In particular, this
  also holds for those where
  $\langle\symp{a'}{b'},\symp{\alpha}{\beta}\rangle_s=0$ as in~\eqref{eq:puncturedStabilizer}, in which case
  $\langle\symp{\mathbf{u}}{\mathbf{v}},
  \symp{\mathbf{u}'}{\mathbf{v}'}\rangle_s=0$. 
  But these $\symp{\mathbf{u}'}{\mathbf{v}'}$ are exactly the elements
  of $S_p^{\symp{\alpha}{\beta}}$, so
  $\symp{\mathbf{u}}{\mathbf{v}}\in
  (S_p^{\symp{\alpha}{\beta}})^\perp$ as claimed.

  For the other inclusion, let $\symp{\mathbf{u}}{\mathbf{v}}\in
  (S_p^{\symp{\alpha}{\beta}})^\perp$. We split the proof in two
  separate cases.  The first case is
  $\ker\sigma_{\symp{\alpha}{\beta}} = S_p$, meaning that every
  element of $S_p$ must be of the form $\symp{a',\mathbf{u}'}{b',\mathbf{v}'}$ with
  $\langle\symp{a'}{b'},\symp{\alpha}{\beta}\rangle_s=0$.  In particular, this also means
  that $\symp{\mathbf{u}'}{\mathbf{v}'}\in
  S_p^{\symp{\alpha}{\beta}}$.  But then
  \begin{align*}
    \langle\symp{\alpha,\mathbf{u}}{\beta,\mathbf{v}}&, \symp{a',\mathbf{u}'}{b',\mathbf{v}'}\rangle_s\\
                                                     &= \langle \symp{\alpha}{\beta}, \symp{a'}{b'}\rangle_s + \langle\symp{\mathbf{u}}{\mathbf{v}}, \symp{\mathbf{u}'}{\mathbf{v}'}\rangle_s\\
    &= 0.
  \end{align*}
  Thus, $\symp{\alpha,\mathbf{u}}{\beta,\mathbf{v}}\in S_p^\perp$ and
  $\symp{\mathbf{u}}{\mathbf{v}}\in (S_p^\perp)^{\symp{\alpha}{\beta}}$ as desired. 

  The second case is $\dim\ker\sigma_{\symp{\alpha}{\beta}}=\dim S_p -1$.
  Here, we may fix any basis of
  $\ker\sigma_{\symp{\alpha}{\beta}}$ and extend it to a basis of
  $S_p$ by adding a single basis element
  $\symp{a',\mathbf{u}'}{b',\mathbf{v}'}$. Importantly, this basis
  element must have $\langle \symp{a'}{b'},
  \symp{\alpha}{\beta}\rangle_s = \delta$ for some non-zero $\delta\in
  \F_p$.  Now, let $\gamma = \langle
  \symp{\mathbf{u}}{\mathbf{v}},
  \symp{\mathbf{u}'}{\mathbf{v}'}\rangle_s$, and define $\symp{a}{b} =
  \gamma\delta^{-1}\symp{\alpha}{\beta}$. We obtain
  \begin{align*}
    \langle\symp{a,\mathbf{u}}{b&,\mathbf{v}}, \symp{a',\mathbf{u}'}{b',\mathbf{v}'}\rangle_s\\
    &= \gamma\delta^{-1}\langle\symp{\alpha}{\beta}, \symp{a'}{b'}\rangle_s
      + \langle\symp{\mathbf{u}}{\mathbf{v}}, \symp{\mathbf{u}'}{\mathbf{v}'} \rangle_s\\
                                            &= \gamma\delta^{-1}(-\delta) + \gamma\\
    &= 0.
  \end{align*}
  The remaining basis elements are symplectically orthogonal to
  $\symp{a,\mathbf{u}}{b,\mathbf{v}}$ by similar arguments as in the
  first case above. 
  This shows that $\symp{a,\mathbf{u}}{b,\mathbf{v}}\in S_p^\perp$,
  and by the choice of $\symp{a}{b}$ this gives
  $\symp{\mathbf{u}}{\mathbf{v}}\in (S_p^\perp)^{\symp{\alpha}{\beta}}$ as required. 
\end{IEEEproof}

Having established that $S_p^{\symp{\alpha}{\beta}}$ is indeed a
stabilizer, and that this form of puncturing behaves nicely with
respect to the symplectic dual, we now turn our attention to the
dimension of the punctured code.
\begin{lem}\label{lem:dimSpP}
  If $d\geq 2$ and $\symp{\alpha}{\beta}\ne\symp{0}{0}$, then $\dim
  S_p^{\symp{\alpha}{\beta}}=\dim S_p-1$.
\end{lem}
\begin{IEEEproof}
  Define $S_p^1=\{\symp{a_1}{b_1}\mid \symp{\vek{a}}{\vek{b}}\in
  S_p\}$, which is restricting $S_p$ to the first coordinate, and
  denote by $\tilde{\pi}_1$ the restriction of $\pi_1$ to $\ker
  \sigma_{\symp{\alpha}{\beta}}$. Thus, we have 
  \begin{equation*}
    \dim S_p^{\symp{\alpha}{\beta}} = \dim\ker\sigma_{\symp{\alpha}{\beta}} - \dim\ker\tilde{\pi}_1.
  \end{equation*}
  Note that due to the condition $\langle
  \symp{a_1}{b_1},\symp{\alpha}{\beta}\rangle_s=0$ on elements
  $\symp{\mathbf{a}}{\mathbf{b}}\in \ker
  \sigma_{\symp{\alpha}{\beta}}$, the kernel of $\tilde{\pi}_1$ can at
  most be one-dimensional. 
  We now consider two cases.

  If $\mathrm{span}(S_p^1)\subseteq
  \mathrm{span}(\{\symp{\alpha}{\beta}\})$, we have
  $\sigma_{\symp{\alpha}{\beta}}(S_p)=\{0\}$, meaning that
  $\ker{\sigma_{\symp{\alpha}{\beta}}}=S_p$. However, this also
  implies that $\symp{\alpha,\vek{0}}{\beta,\vek{0}}\in S_p^\perp$. It
  cannot be in $S_p^\perp\setminus S_p$ due to the assumption on the
  minimum distance, and hence $\symp{\alpha,\vek{0}}{\beta,\vek{0}}\in
  S_p$. Clearly, $\symp{\alpha,\vek{0}}{\beta,\vek{0}}$ is also in
  $\ker \tilde{\pi}_1$, meaning that its kernel is one-dimensional.

  If
  $\mathrm{span}(S_p^1)\not\subseteq\mathrm{span}(\{\symp{\alpha}{\beta}\})$,
  we know that $\dim \ker \sigma_{\symp{\alpha}{\beta}}=\dim
  S_p-1$. Thus, all we need to show is that $\tilde{\pi}_1$ has a
  trivial kernel. 
  Elements of the kernel are $\symp{c\alpha,\vek{0}}{c\beta,\vek{0}}$
  for some $c\in \F_p$. However, we know that there exists an
  $\symp{a}{b}\in S_p^1\notin \mathrm{span}(\{\symp{\alpha}{\beta}\})$
  implying that $\langle\symp{a}{b},\symp{\alpha}{\beta}\rangle_s\neq
  0$. 
  There is a corresponding element $\symp{a,\vek{u}}{b,\vek{v}}\in
  S_p$, and since $S$ is Abelian we have 
  \begin{equation*}
      0=\langle \symp{c\alpha,\vek{0}}{c\beta, \vek{0}},\symp{a,\vek{u}}{b,\vek{v}}\rangle_s=c\langle \symp{\alpha}{\beta},\symp{a}{b}\rangle_s.
  \end{equation*}
    This can only be satisfied for $c=0$, meaning that $\dim \ker \tilde{\pi}_1=0$.
\end{IEEEproof}

\begin{lem}\label{lem:dimSpDual}
  If $d\geq 2$ and $\symp{\alpha}{\beta}\ne\symp{0}{0}$, then $\dim
  (S_p^\perp)^{\symp{\alpha}{\beta}}=\dim S_p^\perp -1$.
\end{lem}
\begin{IEEEproof}
  By Proposition~\ref{prop:dualSpP} and Lemma~\ref{lem:dimSpP}, we obtain
  \begin{align*}
    \dim (S_p^\perp)^{\symp{\alpha}{\beta}} &= \dim (S_p^{\symp{\alpha}{\beta}})^\perp\\
                       &=2(n-1) - \dim S_p^{\symp{\alpha}{\beta}}\\
                       &=2n - \dim S_p -1\\
                       &=\dim S_p^\perp -1
  \end{align*}
  as claimed.
\end{IEEEproof}
In the case where $d=1$, the description of the parameters is less straight-forward because it depends on the choice of $\symp{\alpha}{\beta}$ and on the structure of $S_p$. Later, we will use puncturing to remove an element of weight $1$ from $S_p^{\perp}\setminus S_p$, and in this case we can determine the dimensions.
\begin{lem}\label{lem:dimsFinalPuncture}
  Assume $d=1$ and $\symp{\alpha, \mathbf{0}}{\beta, \mathbf{0}}\in S_p^\perp\setminus S_p$. Then $\dim S_p^{\symp{\alpha}{\beta}} = \dim S_p$ and $\dim (S_p^\perp)^{\symp{\alpha}{\beta}} = \dim S_p^\perp -2$.
\end{lem}
\begin{IEEEproof}
  Since $\symp{\alpha, \mathbf{0}}{\beta, \mathbf{0}}\in S_p^\perp$,
  all elements of $S_p$ must have the form
  $\symp{a,\mathbf{u}}{b,\mathbf{v}}$ for some $a,b$ with
  $\langle\symp{a}{b}, \symp{\alpha}{\beta}\rangle_s=0$.  Hence,
  $\ker\sigma_{\symp{\alpha}{\beta}} = S_p$, and the dimension can
  only be reduced if $\tilde{\pi}_1$ has nontrivial kernel, where
  $\tilde{\pi}_1$ is the same as in the proof of
  Lemma~\ref{lem:dimSpP}. An element is in this kernel if and only if
  it has the form $\symp{c\alpha,\mathbf{0}}{c\beta, \mathbf{0}}$ for
  some $c\in\F_p$. But $c\neq 0$ contradicts
  $\symp{\alpha,\mathbf{0}}{\beta,\mathbf{0}}\in S_p^\perp\setminus
  S_p$, so $\tilde{\pi}_1$ is injective and the dimension is
  preserved.

  The second claim follows by using $\dim S_p^{\symp{\alpha}{\beta}} = \dim S_p$ in calculations identical to the ones in the proof of Lemma~\ref{lem:dimSpDual}.
\end{IEEEproof}
Since $S_p^{\symp{\alpha}{\beta}}$ represents a stabilizer code, it is possible to puncture this code as well. Therefore, we will sometimes use the notation $S_p^{\symp{\vek{\alpha}}{\vek{\beta}}}$ for the code obtained by a sequence of consecutive puncture operations determined by $\symp{\alpha_1}{\beta_1}$, then $\symp{\alpha_2}{\beta_2}$ etc.

\subsection{A stabilizer matrix viewpoint}\label{sec:stabilizerMatrix}

Throughout this section, assume $d\geq 2$ and to simplify notation assume that we puncture the first index with respect to $\symp{\alpha}{\beta}$. To obtain the punctured stabilizer matrix, we cannot simply remove rows that do not satisfy the commutativity
condition on the first index since the span of such rows may contain
vectors where the condition \emph{is} satisfied.  Instead, we first
perform row operations to ensure that only a single row in $G_S$ is removed in this fashion. Note how this corresponds to the results in
Lemmas~\ref{lem:dimSpP} and \ref{lem:dimSpDual}. In this way, the
stabilizer matrix of $S_p$ can be used directly to compute the
stabilizer matrix of $S_p^{\symp{\alpha}{\beta}}$. The row operations carried out are summarized in Algorithm \ref{alg:row_operations}. Note how the row operations in step \ref{step:row_operation}) ensure that all rows except row $i$ satisfies $\langle\symp{a_1^{(j)}}{b_1^{(j)}}, \symp{\alpha}{\beta}\rangle_s=0$. 

We illustrate the proposed puncturing procedure and highlight the benefits from the freedom of choosing $\symp{\alpha}{\beta}$ in the following example.

\begin{algorithm}[t]
\caption{Row operations to obtain stabilizer matrix of punctured code when $d\geq 2$}
  \label{alg:row_operations}
  Let $G=\begin{bmatrix}
      G^{(1)}\\
      \vdots\\
      G^{(n-k)}\\
      \hline
      G^{(n-k+1)}\\
      \vdots\\
      G^{(n+k)}
  \end{bmatrix}=\begin{bmatrix}
      \vek{a}^{(1)}|\vek{b}^{(1)}\\
      \vdots\\
      \vek{a}^{(n-k)}|\vek{b}^{(n-k)}\\
      \hline
      \vek{a}^{(n-k+1)}|\vek{b}^{(n-k+1)}\\
      \vdots\\
      \vek{a}^{(n+k)}|\vek{b}^{(n+k)}
  \end{bmatrix}$ be a centralizer matrix, and let $\symp{\alpha}{\beta}$ be the element we puncture the first column with respect to. 
   \begin{enumerate}
 \item  Let $i$ be the first row satisfying $\delta:=\langle\symp{a^{(i)}_1}{b^{(i)}_1},\symp{\alpha}{\beta}\rangle_s\neq 0$. 
 \item\label{step:row_operation}  For $j=i+1\ldots n+k$: If $\gamma:= \langle\symp{a^{(j)}_1}{b^{(j)}_1},\symp{\alpha}{\beta}\rangle_s\neq 0$, set 
 \begin{alignat*}{5}
 G^{(j)}=G^{(j)}-\gamma \delta^{-1}G^{(i)}
 \end{alignat*}
     \item Remove row $i$ and the first column of $G$. 
  \end{enumerate}
\end{algorithm}

\begin{ex}\label{ex:puncturing}
    We consider a $[\![5,2,2]\!]_3$ code with centralizer matrix
    \begin{align}\label{eq:exam_ex_stab}
      \setlength{\arraycolsep}{4pt}
      \left[
        \begin{array}{*{11}{c}}
        1&0&0&0&0&\vline&0&1&2&2&2\\
        0&1&0&2&0&\vline&2&1&1&1&0\\
        0&0&1&0&1&\vline&1&0&2&2&1\\  
        \hline
        0&0&0&1&0&\vline&2&1&0&0&2\\
        0&0&0&0&1&\vline&2&0&0&0&1\\
        0&0&0&0&0&\vline&0&2&0&2&0\\
        0&0&0&0&0&\vline&0&2&1&2&2
        \end{array}\right].
    \end{align}
    Note that the fifth row of the matrix implies that $S_p^\perp\setminus S_p$ contains the vector
    \begin{align*}
        \setlength{\arraycolsep}{4pt}
      \left[
        \begin{array}{*{11}{c}}
            0&0&0&0&2&\vline&1&0&0&0&2
        \end{array}\right].
    \end{align*}
    This vector will remain if we puncture with respect to
    $\symp{0}{1}$, and once the first index is removed, we are left
    with a vector of symplectic weight one. That is, puncturing with
    respect to $\symp{\alpha}{\beta}=\symp{0}{1}$ decreases the
    distance by one. 

    In an attempt to preserve the distance, we may list all elements
    (up to scalar multiplication) that have symplectic weight $2$ and
    are nonzero on the first pair of entries. These are
    \begin{align*}
      \setlength{\arraycolsep}{4pt}
      \begin{array}{*{11}{c}}
        (1&0&0&1&0&|&2&0&0&0&0),\\
        (1&0&1&0&0&|&2&0&0&0&0),\\
        (0&1&0&0&0&|&1&1&0&0&0),\\
        (0&0&0&0&2&|&1&0&0&0&2).
      \end{array}
    \end{align*}
    Observe that none of these elements, nor their scalar multiples,
    have $\symp{1}{1}$ as their first entry.  Subsequently, puncturing
    with respect to $\symp{1}{1}$ will imply a minimum distance of at
    least $2$ in the punctured code.

    We illustrate how to perform this puncturing from the centralizer matrix in \eqref{eq:exam_ex_stab}. First, perform row
    operations to ensure that all rows except one in the stabilizer
    matrix has a first entry which lies within the span of
    $\symp{1}{1}$ as described in Algorithm~\ref{alg:row_operations}.  Note that the upper part of the centralizer matrix describes $S_p$, and to preserve this property,
    only linear combinations within $S_p$ are allowed. In other words,
    row operations on the first three rows of this example, may only
    involve these three rows.
    
    In the lower part, all linear combinations are allowed as long as
    they do not correspond to elements of $S_p$. That is, no sequence
    of row operations is allowed to completely `eliminate' the original
    element of the extended basis. Thus the matrix after row
    operations is
    \begin{align}\label{eq:exam_ex_stab_row_op}
      \setlength{\arraycolsep}{4pt}
      \left[
        \begin{array}{>{\columncolor{black!15}}cccccc>{\columncolor{black!15}}ccccc}
        \rowcolor{black!15}1&0&0&0&0&\vline&0&1&2&2&2\\
        2&1&0&2&0&\vline&2&0&2&2&1\\  1&0&1&0&1&\vline&1&1&1&1&0\\  
        \hline
        2&0&0&1&0&\vline&2&0&1&1&0\\
        2&0&0&0&1&\vline&2&2&1&1&2\\
        0&0&0&0&0&\vline&0&2&0&2&0\\
        0&0&0&0&0&\vline&0&2&1&2&2
        \end{array}\right],
    \end{align}
    and removing the marked columns and row, will result in a centralizer matrix for the punctured code. The row is removed
    since it is not in the kernel of $\sigma_{\symp{1}{1}}$, and the
    columns are removed by $\pi_1$.
    
    It can be shown that this is in fact a $[\![4,2,2]\!]_3$-code,
    meaning that we have only lowered $n$ without changing $k$ and
    $d$. This code meets the quantum
    Singleton-bound~\cite{KnillLaflamme,rains}, and is therefore
    optimal.
\end{ex}

\subsection{Idea for constructing good codes}\label{sec:Idea_to_good_codes}
Having established the connection between puncturing and its effect on
the stabilizer matrix, we now describe how this may be used to search
for codes with good parameters according to \cite{Grassl:codetables}.

\begin{prop}
    Consider an $[\![n,k,d]\!]_p$ stabilizer code defined from $S_p$. Let $I$ be a set of indices with $|I|=n-1$, and assume that $\symp{\alpha}{\beta}\notin \pi_{I}(S_p^\perp\setminus S_p)$ and $\symp{\alpha}{\beta}\neq \symp{0}{0}$. Then the punctured code on index $\{1,2,\ldots,n\}\setminus I$ according to $\symp{\alpha}{\beta}$ has minimum distance $d'\geq d$.
\end{prop}
\begin{IEEEproof}
    For $d=1$ the result is trivial. Now, let $d\geq 2$ and to simplify the notation assume that $I=\{2,3,\ldots,n\}$. From Lemmas~\ref{lem:dimSpP} and \ref{lem:dimSpDual}, we need to consider elements $\symp{a,\mathbf{u}}{b,\mathbf{v}}\in
S_p^\perp\setminus S_p$ of symplectic weight $d$. If
$\symp{a}{b}=\symp{0}{0}$ the weight will still be $d$ after
puncturing. Now consider the case where $(a|b)\neq (0|0)$. Since
$\symp{\alpha}{\beta}$ never appears as a first entry, we have
$\symp{a}{b}\neq c\symp{\alpha}{\beta}$ for any $c\in
\F_p$. This implies that for any
$\symp{a',\mathbf{u}'}{b',\mathbf{v}'}\in
\ker\sigma_{\symp{\alpha}{\beta}}\subseteq S_p$, we have
$\langle\symp{a}{b},\symp{a'}{b'}\rangle_s\neq 0$ by Lemma \ref{lem:puncture}. By linearity of the
symplectic form we conclude that
$\langle\symp{\mathbf{u}}{\mathbf{v}},\symp{\mathbf{u}'}{\mathbf{v}'}\rangle_s\neq
0$.  But $\symp{\mathbf{u}'}{\mathbf{v}'}\in
\pi_1(\ker\sigma_{\symp{\alpha}{\beta}})=S_p^{\symp{\alpha}{\beta}}$,
so it must necessarily be the case that
$\symp{\mathbf{u}}{\mathbf{v}}\notin
(S_p^{\symp{\alpha}{\beta}})^\perp$. In other words, all the elements
$\symp{a,\mathbf{u}}{b,\mathbf{v}}\in S_p^\perp\setminus S_p$ with
$\symp{a}{b}\neq \symp{0}{0}$ of symplectic weight $d$ are removed
during puncturing, and hence $d'\geq d$.
\end{IEEEproof}

This gives us a strategy to find codes with minimum distance improving the lower bound $d'\geq d-1$. Assume that we are able to list all elements of $S_p^\perp\setminus S_p$ of symplectic weight $d$. With this list, we may search for an index where the entries of the minimum weight elements do not exhaust all possible values. If we let $\symp{\alpha}{\beta}$ be such an element, then puncturing with respect to $\symp{\alpha}{\beta}$ on this position will give us an $[\![n-1,k,d']\!]_p$-code, where $d'\geq d$.

We remark that this strategy of puncturing codes is very similar to
the idea of hitting sets in classical codes, see
\cite{GrasslWhite2004}, and we generalize this idea when applying $t$ successive puncturings. For classical codes, puncturing at $t$
positions does not decrease the minimum distance by $t$ if every
minimum weight word has at least one zero among those positions.  For
quantum codes, $\symp{a_i}{b_i}$ may be non-zero if $\langle
\symp{a_i}{b_i},\symp{\alpha_i}{\beta_i}\rangle_s \ne0$.  This is
equivalent to $\symp{a_i}{b_i}\ne c_i\symp{\alpha_i}{\beta_i}$ for any
$c_i\in\F_p$.  Then the corresponding vector is not in the
kernel of $\sigma_{\symp{\alpha_i}{\beta_i}}$ and is removed.  Let
$\mathcal{M}$ denote the set of all minimum weight
codewords of $S_p^\perp\setminus S_p$.  For a fixed set of indices
$I\subseteq\{1,\ldots,n\}$ with $|I|=t$, we consider the tuples
$\mathcal{M}_I=\{ (v_i)_{i\in I}\colon \vek{v}\in\mathcal{M}\}$.  When
$v_i=\symp{0}{0}$ for some $i\in I$, puncturing at $I$ results in a
vector of weight larger than $d-t$.  Hence it is sufficient to
consider the $t$-tuples of symplectic weight $t$, which we denote by
$\mathcal{M}_{I,t}$.
We form the set
\begin{align*}
  \mathcal{M}^*_{I,t}=\{ \symp{c_1 a_1,\ldots,c_t a_t}{c_1 b_1,\ldots,c_t b_t}
   \colon\vek{c}\in(\F_p^*)^t, \symp{\vek{a}}{\vek{b}}\in\mathcal{M}_{I,t}\},
\end{align*}
where $\F_p^*=\F_p\setminus\{0\}$.  
\begin{prop}\label{prop:improved_dist}
    If
$|\mathcal{M}^*_{I,t}|<(p^2-1)^t$, then there exists
$\symp{\vek{\alpha}}{\vek{\beta}}$ such that puncturing with respect to this $\symp{\vek{\alpha}}{\vek{\beta}}$ and the indices of $I$ gives an $[\![n-t,k,d']\!]_p$ code with $d'> d-t$.
\end{prop}
\begin{IEEEproof}
    If $|\mathcal{M}^*_{I,t}|<(p^2-1)^t$ there exists
$\symp{\vek{\alpha}}{\vek{\beta}}\notin\mathcal{M}_I$ such that
$\symp{\alpha_i}{\beta_i}\ne\symp{0}{0}$ for $1\le i\le t$, and for
every $\symp{\vek{a}}{\vek{b}}\in\mathcal{M}_{I,t}$ there exists an
index $i$ such that
$\langle\symp{a_i}{b_i},\symp{\alpha_i}{\beta_i}\rangle_s\ne 0$.
Hence, puncturing removes all minimum weight words for which $I$ is
contained in their support. Therefore, the punctured code has at least
minimum distance $d-t+1$.
\end{IEEEproof}

\subsection{Relation to existing puncturing descriptions}\label{sec:relationGrassl}
As mentioned in the introduction, a puncturing procedure for general
quantum codes has already been described in~\cite[Sec.\,4]{Grassl21}
and for qubit stabilizer codes in the language of additive
$\F_4$-codes in \cite[Thm.\,6]{CodesF4}. 
Given a quantum code $[\![n,k,d]\!]_q$ with $d>1$, these procedures yield a quantum code $[\![n-1,k,\ge d-1]\!]_q$.
We now describe how these procedures relate to our description.  In \cite{CodesF4}, the bijective map
\begin{align*}
    \phi\colon \F_2^2&\to \F_4\\
    \symp{a}{b}&\mapsto\omega a+\bar{\omega}b
\end{align*}
is introduced, where $\omega$ is an element of the extension
field. This allows to interpret $S_p^\perp$ and $S_p$ as additive
codes over $\F_4$. Puncturing is then defined as taking the
elements $(a,\vek{u}|b,\vek{v})$ from $S_p^\perp$ and $S_p$ having $\phi((a|b))=0$ or $\phi((a|b))=1$ as
their first entry, and afterwards removing this entry. Note, however,
that $\phi(\symp{a}{b})=0$ implies that $\symp{a}{b}=\symp{0}{0}$, and
$\phi(\symp{a}{b})=1$ implies that $\symp{a}{b}=\symp{1}{1}$. Thus,
this is a special case of our puncture procedure, choosing
$\symp{\alpha}{\beta}=\symp{1}{1}$ which is described in the case
$p=2$. We have already seen the benefits of our generalization in Example \ref{ex:puncturing} where the freedom of choosing $\symp{\alpha}{\beta}$ leads to improved parameters.

To relate our results to \cite{Grassl21} we will consider the case
$q=p>2$ prime. The binary setting is a special case that needs
slightly modified arguments as was also discussed in
Section~\ref{sec:stabilizer-codes} in relation to
$\mathcal{G}_n$. Including this special case would reduce the clarity
of the overall idea, so we will leave it out.

In \cite[Sec. 4]{Grassl21}, puncturing a stabilizer code with $d>1$ is
carried out by fixing one of the qudits to a specific state.  More
precisely, to puncture the first coordinate we fix a pure state
$\ket{\psi}$ and apply the projection $P_\psi =
\ket{\psi}\bra{\psi}\otimes I^{\otimes (n-1)}$ to the code.  If
$P_\psi\ket{c}=\vek{0}$ for some state $\ket{c}$ in the code, then the
Knill-Laflamme conditions \cite{KnillLaflamme} imply that
$P_\psi\ket{c}=\vek{0}$ for all states in the code. Choosing $\ket{\psi}$
such that the code is not in the kernel of $P_{\psi}$, each basis
state $\ket{c_i}$ of $\mathcal{C}_S$ is mapped to $\ket{\psi}\otimes\ket{c_i'}$,
and $\{\ket{c_i'}\}_{i\in\F_2^k}$ spans a new quantum code.

Assume now that the code space is stabilized by $S$, and write each
element $M\in S$ as $M=A\otimes B$, where $A$ acts on a single qudit
and $B$ on $n-1$ qudits.\footnote{Because of the global phase, this
decomposition is not unique. But importantly, the specific choice does
not affect the following arguments.}  The assumption $d>1$ implies
$A\notin\langle\omega I\rangle$ for at least one element. Otherwise,
$A'\otimes I^{\otimes (n-1)}$ would be in $C(S)\setminus S$ for every
non-trivial $A'\in\mathcal{G}_1$. Hence, $d=1$ in that case.

Thus, we may choose some $M^\ast=A^\ast \otimes B^\ast \in S$ with
$A^\ast \notin\langle\omega I\rangle$. Applying the
isomorphism~\eqref{eq:iso_to_field} to $A^\ast$ yields some
$\symp{\alpha}{\beta}\in\F_p^{2}$.  Using the procedure from
Section~\ref{sec:punctureSymplectic} to puncture $S$ according to
$\symp{\alpha}{\beta}$ now corresponds to finding those $M=A\otimes
B\in S$ where $A$ and $A^\ast$ commute, and then letting $B$ be part
of the punctured stabilizer.

To describe such operators, we will utilize the structure of their
eigenspaces as indicated in the following lemma.
\begin{lem}\label{lem:eigenspacesAreLines}
  Let $p>2$. If $A\in\mathcal{G}_1\setminus \langle\omega I\rangle$,
  then $A$ has $\omega^k$ as an eigenvalue for all
  $k\in\{0,1,\ldots,p-1\}$. In addition, every eigenspace is
  one-dimensional.
\end{lem}
\begin{IEEEproof}
  Since $A$ is unitary, all of its eigenvalues $\lambda\in\mathbb{C}$
  have modulus $|\lambda| = 1$, see
  e.\,g., \cite[p.~583]{LangAlgebra}. Furthermore, $A^p=I$ implies
  $\lambda^p=1$, i.\,e., all eigenvalues are $p$-th roots of unity.
  To prove that the eigenvalues are exactly the $p$-th roots of unity,
  write $A=\omega^c X(a)Z(b)$ where we note that $(a,b)\neq (0,0)$
  since this would contradict $A\notin\langle\omega I\rangle$.  Now,
  consider the element $A'=X(a')Z(b')$. We have $AA'=\omega^{a'b-ab'}
  A'A$, and $(a,b)\neq (0,0)$ implies that $a',b'$ can be chosen such
  that $a'b-ab'=1$. In other words, we can choose $A'$ such that
  \begin{equation*}
    A(A')^k = \omega^k (A')^kA
  \end{equation*}
  for any $k\in\{0,1,\ldots,p-1\}$. If we now let $\ket{\varphi}$ be
  any $\lambda$-eigenvector of $A$, then
  \begin{equation*}
    A (A')^k\ket{\varphi} = \omega^k (A')^k A\ket{\varphi} =\lambda\omega^k (A')^k\ket{\varphi},
  \end{equation*}
  meaning that $(A')^k\ket{\varphi}$ is a
  $\lambda\omega^k$-eigenvector of $A$. Thus, we get exactly all the
  $p$-th roots of unity $\{1,\omega,\ldots,\omega^{p-1}\}$ as
  eigenvalues of $A$.  The sum of the corresponding eigenspace
  dimensions is at most $p$, yielding the final remark of the lemma.
\end{IEEEproof}
Now, recall the projection $P_\psi$, and let $\ket{\psi}$ be an
eigenvector of $A^\ast$, such that the puncturing of~\cite{Grassl21}
fixes the first qudit to be $\ket{\psi}$.  Consider $M=A\otimes B$ as
before, and assume that $A$ commutes with $A^\ast$. Both operators are
diagonalizable, and their commutativity implies that they are
simultaneously diagonalizable, see
e.\,g.,~\cite[Thm. 1.3.21]{HJ13}. That is, there exists a basis of
$\mathbb{C}^p$ consisting of common eigenvectors of both $A^\ast$ and
$A$. But by Lemma~\ref{lem:eigenspacesAreLines} the eigenspaces of
$A^\ast$ are one-dimensional, implying $\ket{\psi}$ must also be an
eigenvector of $A$.

So for the puncturing of Section~\ref{sec:punctureSymplectic}, we pick those $A\in\mathcal{G}_n$ that
have $\ket{\psi}$ as an eigenvector.  Clearly, $\ket{\psi}$ is a $1$-eigenvector of $\ket{\psi}\bra{\psi}$, and the orthogonal complement of the corresponding $1$-eigenspace forms the $0$-eigenspace of $\ket{\psi}\bra{\psi}$.
    Furthermore, since $A$ is unitary, the Spectral Theorem ensures that its eigenspaces are orthogonal. Combined with Lemma~\ref{lem:eigenspacesAreLines}, this implies that the remaining eigenvectors of $A$ (that is, those apart from $\ket{\psi}$) form a basis of the $0$-eigenspace of $\ket{\psi}\bra{\psi}$.
    Hence, $A$ and $\ket{\psi}\bra{\psi}$ are simultaneously diagonalizable, and they commute.
    Subsequently, any $M=A\otimes B\in S$ where $A$ has $\ket{\psi}$ as an eigenvector
    must commute with $P_\psi = \ket{\psi}\bra{\psi}\otimes I^{\otimes (n-1)}$.
    Consider any $M=A\otimes B\in S$ with these properties (in
particular, $M^\ast$ is one such operator).  If we denote by $\lambda$
the eigenvalue of $\ket{\psi}$ and use the definition of a
stabilizer~\eqref{eq:stabilizerDef}, we obtain
\begin{equation*}
  M P_\psi\ket{c_i} = P_\psi\ket{c_i} = \ket{\psi}\otimes\ket{c_i'},
\end{equation*}
and
\begin{align*}
  M P_\psi\ket{c_i} &= (I\otimes B)(A\ket{\psi}\bra{\psi} \otimes I^{\otimes (n-1)})\ket{c_i}\\
                     &= (I\otimes B)(\lambda \ket{\psi}\otimes\ket{c_i'})\\
                     &=\ket{\psi} \otimes (\lambda B)\ket{c_i'},
\end{align*}
where we ignore normalization factors. Combining these, we see both that
\begin{enumerate*}[label=(\alph*), itemjoin={{, and }}]
  \item $M$ stabilizes the projected code states $P_\psi\ket{c_i}$
  \item $\lambda B$ stabilizes the new, punctured code states $\{\ket{c_i'}\}$.
\end{enumerate*}
Additionally, since $M=A\otimes B$ and $M^\ast=A^\ast\otimes B^\ast$
are in $S$, they must commute. Combined with the fact that $A$ and
$A^\ast$ commute, this implies that $B$ and $B^\ast$ must also
commute.  All together, these observations show that the set of
$\lambda B$ form a new stabilizer with code space spanned by
$\{\ket{c_i'}\}$.

As a final remark, note that those $A$ that do not commute with
$A^\ast$ cannot have $\ket{\psi}$ as an eigenvector. Hence, such $M$
cannot stabilize the projected states $\ket{\psi}\otimes \ket{c_i'}$.

We would also like to note that if one chooses a projection $P_\psi$
such that the state $\ket{\psi}$ is not an eigenstate of any
non-trivial $A\in\mathcal{G}_n$, i.\,e., $\ket{\psi}$ is not a
`stabilizer state', then the code resulting from puncturing using
$P_\psi$ is in general not a stabilizer code, even if one started with
a stabilizer code.
Therefore, our construction can be seen as a special case of the construction in \cite{Grassl21}, but with the guarantee that the punctured code is always a stabilizer code. In addition, our construction gives an explicit description of the stabilizers. Moreover, the procedure in \cite{Grassl21} does not make use of the freedom in choosing the state $\ket{\psi}$.

For the special case of CSS codes it has been shown in \cite{Grassl_2023} that a CSS code $[\![n,k,d]\!]_q$ with $n>2$ and $d>1$ can always be punctured at two positions in such a way that the minimum distance is reduced by at most one, i.\,e., one obtains a code $[\![n-2,k,d-1]\!]_q$.
Related, in \cite{LaGuardia2014} it has been shown that puncturing of asymmetric quantum codes from the CSS construction reduces only one of the two relevant minimum distances.

\subsection{Prime power dimensions}\label{sec:nonprime}
Next we will briefly describe puncturing of qudit stabilizer codes for
the case $q=p^r$, $r>1$. 
Recall that in the prime case,
$S_p^{\symp{\alpha}{\beta}}=\pi_1(\ker\sigma_{\symp{\alpha}{\beta}})$
where
\begin{align}\label{eq:sigma_prime}
  \sigma_{\symp{\alpha}{\beta}}\symp{\vek{a}}{\vek{b}} =
    \langle \symp{a_1}{b_1},\symp{\alpha}{\beta}\rangle_s
   = \tr(b_1\alpha-\beta a_1)= b_1\alpha-\beta a_1.
\end{align}
Hence the kernel consists of all elements $\symp{\vek{a}}{\vek{b}}$
with $\symp{a_1}{b_1}=c\symp{\alpha}{{\beta}}$ for some $c\in\F_p$.
In the non-prime case, we cannot omit the trace in
\eqref{eq:sigma_prime}. Then the kernel contains more elements. Also,
the eigenspaces of the operator $A=X(\alpha)Z(\beta)$ have dimension
$q/p=p^{r-1}$. 

Instead of just considering a single $\symp{\alpha}{\beta}$ for a
fixed position, we choose $r$ pairs
$\symp{\alpha^{(j)}}{\beta^{(j)}}$, $1\le j\le r$ such that the trace
symplectic product among all these pairs is zero, i.\,e.,
\begin{align*}
    \langle \symp{\alpha^{(j)}}{\beta^{(j)}},\symp{\alpha^{(j')}}{\beta^{(j')}}\rangle_s
   = \tr\left(\beta^{(j')}\alpha^{(j)}-\beta^{(j)}\alpha^{(j')}\right) = 0\quad\text{for all $1\le j<j'\le r$,}
\end{align*}
and such that they span an $r$-dimensional $\F_p$ vector space, i.\,e.,
they are additively independent. This corresponds to choosing a maximal Abelian
subgroup of $\mathcal{G}_1$ which has one-dimensional eigenspaces.

We replace the map defined in \eqref{eq:symplectic_sigma} by the map
\begin{align}
  \widetilde{\sigma}_{\symp{\alpha^{(j)}}{\beta^{(j)}}}\symp{\vek{a}}{\vek{b}}
  = \left(
  \langle \symp{a_1}{b_1},\symp{\alpha^{(1)}}{\beta^{(1)}}\rangle_s,
  \ldots,
  \langle \symp{a_1}{b_1},\symp{\alpha^{(r)}}{\beta^{(r)}}\rangle_s
 \right)\in\F_p^r.\label{eq:sigmatilde}
\end{align}
The image of the map $\widetilde{\sigma}$ in \eqref{eq:sigmatilde} is a vector over $\mathbb{F}_p$ with $r$ components.
The $j$-th component is zero if and only if the corresponding element of the stabilizer $S$ commutes with the operator $X(\alpha^{(j)})Z(\beta^{(j)})\otimes I^{\otimes (n-1)}$. For an element to be in the kernel, all $r$ components have to be zero. 

Finally, puncturing a qudit code at the first position using the set
$\{\symp{\alpha^{(j)}}{\beta^{(j)}}|j=1,\ldots,r\}$ is given by
\begin{align*}
  S_q^{\symp{\alpha^{(j)}}{\beta^{(j)}}}=\pi_1(\ker\widetilde{\sigma}_{\symp{\alpha^{(j)}}{\beta^{(j)}}}).
\end{align*}
Like in the prime case, we have
$S_q^{\symp{\alpha^{(j)}}{\beta^{(j)}}}\subseteq(S_q^{\symp{\alpha^{(j)}}{\beta^{(j)}}})^\perp$,
and starting from a quantum code $[\![n,k,d]\!]_q$ with $d>1$, puncturing yields
a code $[\![n-1,k,d']\!]_q$ with $d'\ge d-1$.

\subsection{Shortening the stabilizer}\label{sec:shortening}
We will now discuss the role of $\sigma_{\symp{\alpha}{\beta}}$
and the restriction to $\ker \sigma_{\symp{\alpha}{\beta}}$ in the
puncturing procedure, again restricting to the prime case for simplicity.

For classical linear codes, puncturing of a code is related to shortening the dual code.
Let $S_p'$ be the shortened code of $S_p$, i.e., selecting those vectors with
$\symp{a_i}{b_i}=\symp{0}{0}$ in $S_p$. As already mentioned, this corresponds to puncturing the dual code, i.\,e., 
$(S_p')^\perp=\pi_1(S_p^\perp)$. It follows that, in general, one obtains a stabilizer code
with parameters $[\![n-1,k+1,d']\!]_p$.

When the centralizer $C(S)$ corresponding to $S_p^\perp$ does not
contain a non-scalar element of weight strictly smaller than $d$,
i.\,e., if the code is pure, then the resulting code has minimum
distance $d'\ge d-1$. However, if there is a non-scalar element of
weight $\tilde{d}<d$ which is non-trivial on the first qudit, then
$C(S')$ for the new code will contain an element of weight
$\tilde{d}-1<d-1$, which is not contained in the stabilizer $S'$ of
the new code. Hence, in general there is no lower bound on $d'$ just
in terms of $d$ and the number $t$ of puncturings; see also
\cite[Theorem 6 b)]{CodesF4} and \cite[Lemma 71]{Nonbinary_Stabilizer}.

In our puncturing procedure, instead of selecting only those vectors in $S_p$ with $(a_i|b_i)=(0|0)$ before removing one coordinate, we select the vectors in the kernel of $\sigma_{(\alpha|\beta)}$. Then the dimension of the resulting classical code is larger, and the dimension of the corresponding quantum code does not increase. At the same time, the procedure guarantees that the minimum distance is never reduced by more than one.

\section{Quantum Griesmer bound}\label{sec:quant-griesm-bound}
Puncturing is the main tool used for proving the so-called Griesmer
bound in classical error correction. We show that we can generalize
the arguments such that we can prove a Griesmer bound for stabilizer
codes.
Since our quantum analog follows the same proof strategy, the reader may want to recall the classical Griesmer bound \cite{Griesmer60,huffman2003fundamentals}.
\begin{thm}\label{thm:Griesmer_quantum}
  Let $p$ be a prime and consider a stabilizer code with parameters $[\![n,k,d]\!]_p$. The parameters satisfy
  \begin{equation*}
    n\geq \sum_{i=0}^{k-1} \left\lceil \frac{d}{p^i} \right\rceil.
    \end{equation*}
\end{thm}

\begin{IEEEproof}
  Let $S$ be a stabilizer for a code with parameters
  $[\![n,k,d]\!]_p$, and let
  $\symp{\vek{\alpha},\vek{0}}{\vek{\beta},\vek{0}}\in
  S_p^\perp\setminus S_p$ be a vector of minimum symplectic
  weight. Successively puncturing at the first positions according to
  $\symp{\vek{\alpha}}{\vek{\beta}}$, i.\,e., using
  $\symp{\alpha_i}{\beta_i}$ in each step, yields an
  $[\![n-d,k-1,d']\!]_p$ stabilizer code according to Lemmas \ref{lem:dimSpP}, \ref{lem:dimSpDual}, and \ref{lem:dimsFinalPuncture}. We have to show that $d'\geq
  \lceil d/p\rceil$.

  The vectors in $(S_p^{\symp{\vek{\alpha}}{\vek{\beta}}})^\perp\setminus
  S_p^{\symp{\vek{\alpha}}{\vek{\beta}}}$ are of the
  form $\symp{\vek{u}}{\vek{v}}\in \F_p^{2(n-d)}$. These vectors
  are the image of some
  $(\vek{u}'|\vek{v}')=\symp{\vek{a},\vek{u}}{\vek{b},\vek{v}}\in
  \F_p^{2n}$, where for $1\le i\le d$,
  $\symp{a_i}{b_i}=c_i\symp{\alpha_i}{\beta_i}$ for some $c_i\in
  \F_p$, due to Lemma \ref{lem:puncture}.  We show that
  $w_s\symp{\vek{u}}{\vek{v}}$ is at least $d/p$, when
  $\symp{\vek{u}}{\vek{v}}\neq \symp{\vek{0}}{\vek{0}}$.
  
  Let $\tilde{c}$ be the value of the $c_i$
  that occurs most frequently. That means that $\tilde{c}$ occurs at
  least $\lceil \frac{d}{p}\rceil$ times. Note that
  $\symp{\vek{a},\vek{u}}{\vek{b},\vek{v}} -
  \tilde{c}\symp{\vek{\alpha},\vek{0}}{\vek{\beta},\vek{0}} \in
  S_p^\perp\setminus S_p$, since if we consider the first $d$ entries
  with $1\leq i\leq d$, we have
  \begin{align*}
      (a_i-\tilde{c}\alpha_i|b_i-\tilde{c}\beta_i)&=(c_i\alpha_i-\tilde{c}\alpha_i|c_i\beta_i-\tilde{c}\beta_i)\\
      &=((c_i-\tilde{c})\alpha_i|(c_i-\tilde{c}_i)\beta_i).
  \end{align*}
  Thus, applying the puncture procedure to
  $\symp{\vek{a},\vek{u}}{\vek{b},\vek{v}} -
  \tilde{c}\symp{\vek{\alpha},\vek{0}}{\vek{\beta},\vek{0}}$, will
  give us $\symp{\vek{u}}{\vek{v}}$ which is assumed to be a nonzero
  vector in $(S_p^{\symp{\vek{\alpha}}{\vek{\beta}}})^\perp\setminus
  S_p^{\symp{\vek{\alpha}}{\vek{\beta}}}$, and hence $\symp{\vek{a},\vek{u}}{\vek{b},\vek{v}}
  - \tilde{c}\symp{\vek{\alpha},\vek{0}}{\vek{\beta},\vek{0}}$ cannot
  be in $S_p$.

  In that case
  \begin{align*}
    d&\leq w_s\bigl(\symp{\vek{a},\vek{u}}{\vek{b},\vek{v}} - \tilde{c}\symp{\vek{\alpha},\vek{0}}{\vek{\beta},\vek{0}}\bigr)\\
    &=w_s(\vek{a}-\tilde{c}\vek{\alpha}|\vek{b}-\tilde{c}\vek{\beta}) + w_s(\vek{u}|\vek{v})\\
    &\leq \frac{p-1}{p}d + w_s(\vek{u}|\vek{v}),
  \end{align*}
showing that $w_s(\vek{u}|\vek{v})\geq d/p$. 

  Repeating the argument on the punctured code, implies that the Griesmer bound follows by induction on $k$.
\end{IEEEproof}

\section{Quantifying the loss of minimum distance}\label{sec:random-punct}
In Sections~\ref{sec:Idea_to_good_codes} and
\ref{sec:quant-griesm-bound}, we used knowledge of the minimum weight
elements of $S_p^\perp\setminus S_p$ to estimate the distance $d'$ of
the punctured quantum code.  In general, this knowledge may not be
available, in which case we have to settle for the lower bound $d'\geq
d-t$ for $t$ successive puncturings. However, for special types of
codes, for instance CSS codes, it has been shown that it is possible
to design the puncturing such that $d'>d-t$, see \cite{Grassl_2023}.


In this section, we consider two codes to illustrate in a quantitative way 
whether the bound $d'\geq
d-t$ typically matches the parameters obtained via puncturing, or if
the bound turns out to be too pessimistic in many cases. 
We consider the two codes as `mother' codes, and then
compute all the codes that can be obtained by performing $t$ puncturings.
For each of the resulting codes, we compute its minimum distance
$d'$ and compare this to the bound by defining
\begin{equation}\label{eq:Delta}
  \Delta = d'-(d-t).
\end{equation}
In this way, a large value of $\Delta$ indicates codes of better
parameters than guaranteed by the bound.

\begin{figure}[hbt]
  \centering
  \includegraphics{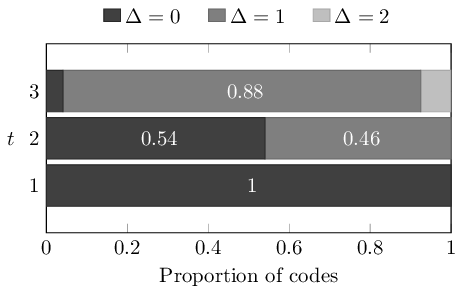}
    \raisebox{5.35ex}{$\def\arraystretch{1.0}\doublerulesep0.5\doublerulesep
  \begin{array}[b]{|c||r||r|r|r|}
    \hline
      t & \text{\#codes} & \Delta=0& \Delta=1& \Delta=2\\
      \hline
      &&&\\[-3.7ex]
      \hline
    3  & 29120 & 1216 & 25740 & 2164\\\hline
    2  & 1680 & 908 & 772&\\\hline
    1  & 60 & 60&&\\\hline
  \end{array}$}

  \caption{Distribution of $\Delta$ as defined in~\eqref{eq:Delta} for the $[\![15,3,5]\!]_3$ code in Example~\ref{ex:trinaryCode}. The unlabeled proportions for $t=3$ are $0.04$ and $0.07$ for $\Delta=0$ and $\Delta=2$, respectively.}
  \label{fig:trinaryCode}
\end{figure}

\begin{figure}[hbt]
  \centering
  \includegraphics{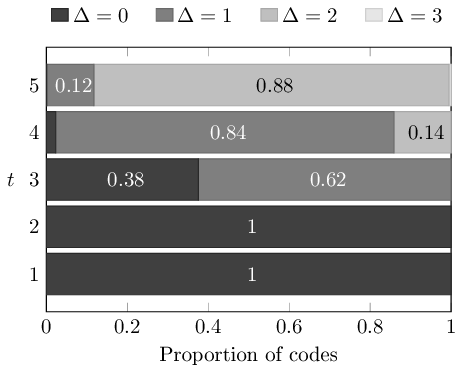}
  \raisebox{5.35ex}{$\def\arraystretch{1.0}\doublerulesep0.5\doublerulesep
  \begin{array}[b]{|c||r||r|r|r|r|}
    \hline
      t & \text{\#codes} & \Delta=0& \Delta=1& \Delta=2& \Delta=3\\
      \hline
      &&&&&\\[-3.7ex]
      \hline
    5  & 235467 & 216 & 27409 & 206604 & 1238\\\hline
    4  & 23085 & 540 & 19285 & 3260&\\\hline
    3  & 1728 & 649 & 1079&&\\\hline
    2  & 90 & 90&&&\\\hline
    1  & 3 & 3&&&\\\hline
  \end{array}$}

  \caption{Distribution of $\Delta$ as defined in~\eqref{eq:Delta} for
    the $[\![21,5,6]\!]_2$ code in Example~\ref{ex:binaryCode}.  For
    $t=4$ and $\Delta=0$, the unlabeled proportion is $2.34\cdot
    10^{-2}$. The unlabeled, and virtually imperceptible, proportions
    for $t=5$ are $9.17\cdot 10^{-4}$ ($216$ out of $235467$) and $5.26\cdot 10^{-3}$ ($1238$
    out of $235467$) for $\Delta=0$ and $\Delta=3$, respectively.}
  \label{fig:binaryCode}
\end{figure}

\begin{ex}\label{ex:trinaryCode}
  We consider the $[\![15,3,5]\!]_3$-code from
  \cite{Grassl:codetables} defined by the stabilizer matrix given
  in~\eqref{eq:bigStabilizer}.  Puncturing this $1$, $2$, and $3$
  times gives $60$, $1680$, and $29120$ different codes,
  respectively. Note that these are the unique codes obtained;
  multiple choices of $\symp{\mathbf{\alpha}}{\mathbf{\beta}}$ may
  lead to the same punctured code.  For example, puncturing with
  respect to $\symp{\alpha}{\beta}$ and $\symp{c\alpha}{c\beta}$ gives
  the same result for any non-zero $c\in\F_p$.  Hence the
  number of different codes when puncturing at $t$ positions is at
  most $\binom{n}{t}(p+1)^t$ for $q=p$ prime. This bound is achieved
  in this case.  The resulting distributions of minimum distances are
  reported in Figure~\ref{fig:trinaryCode}.  As is evident, the bound
  gives the true value after every single puncturing, but becomes less
  and less accurate, as the number of puncturing steps increases.  In
  some cases, three puncturings even resulted in
  $[\![12,3,4]\!]_3$ codes (those with $\Delta=2$). These parameters
  match the best known ones from \cite{Grassl:codetables}.
\end{ex}
\begin{ex}\label{ex:binaryCode}
  The tables of~\cite{Grassl:codetables} list a cyclic
  $[\![21,5,6]\!]_2$ code defined by~\eqref{eq:bigStabilizerBinary},
  and this code is optimal.  We have performed $t=5$ successive
  puncturings of this code which give the minimum distances
  illustrated in Figure~\ref{fig:binaryCode}.
  Due to the cyclic permutation automorphism of the code it is
  sufficient to consider only one representative of each orbit of the
  $t$-subsets with respect to the cyclic group. This reduces the
  number of puncturings to be considered essentially by a factor $n=21$.
  As in Example~\ref{ex:trinaryCode}, the lower bound matches the true
  value for few puncturings, but typically underestimates the minimum
  distance as the number of puncturings increase.  For $t=5$, we
  found a total of $1238$ codes (out of $235467$) that gave $\Delta =
  3$. These have parameters $[\![16,5,4]\!]$, matching the best known
  codes from~\cite{Grassl:codetables}.
\end{ex}

While these are just two examples with a limited number of
puncturings, they do indicate that codes obtained through repeated
puncturing are somewhat likely to have a higher minimum distance than
the worst-case of $d'=d-t$.

\begin{figure*}[hbt]
  \centering
  \small
  \begin{equation}\label{eq:bigStabilizer}
    \left[
      \begin{array}{*{15}{@{\hspace{.7em}}c}|*{15}{c@{\hspace{.7em}}}}
        1 & 0 & 0 & 0 & 0 & 0 & 2 & 1 & 2 & 2 & 0 & 2 & 2 & 1 & 2 & 0 & 0 & 0 & 0 & 0 & 0 & 0 & 1 & 0 & 1 & 1 & 0 & 2 & 0 & 0\\
        0 & 0 & 0 & 0 & 0 & 0 & 0 & 1 & 0 & 1 & 1 & 0 & 2 & 0 & 0 & 1 & 0 & 0 & 0 & 0 & 0 & 2 & 2 & 2 & 0 & 1 & 2 & 1 & 1 & 2\\
        0 & 1 & 0 & 0 & 0 & 0 & 2 & 2 & 1 & 0 & 0 & 0 & 0 & 2 & 2 & 0 & 0 & 0 & 0 & 0 & 0 & 1 & 2 & 2 & 1 & 2 & 0 & 1 & 1 & 2\\
        0 & 0 & 0 & 0 & 0 & 0 & 1 & 2 & 2 & 1 & 2 & 0 & 1 & 1 & 2 & 0 & 1 & 0 & 0 & 0 & 0 & 0 & 1 & 0 & 1 & 2 & 0 & 1 & 0 & 1\\
        0 & 0 & 1 & 0 & 0 & 0 & 1 & 1 & 2 & 2 & 0 & 0 & 2 & 2 & 1 & 0 & 0 & 0 & 0 & 0 & 0 & 1 & 0 & 2 & 0 & 0 & 1 & 1 & 2 & 2\\
        0 & 0 & 0 & 0 & 0 & 0 & 1 & 0 & 2 & 0 & 0 & 1 & 1 & 2 & 2 & 0 & 0 & 1 & 0 & 0 & 0 & 2 & 1 & 1 & 2 & 0 & 1 & 0 & 1 & 0\\
        0 & 0 & 0 & 1 & 0 & 0 & 0 & 1 & 2 & 0 & 0 & 0 & 0 & 2 & 2 & 0 & 0 & 0 & 0 & 0 & 0 & 2 & 2 & 2 & 2 & 0 & 2 & 1 & 0 & 1\\
        0 & 0 & 0 & 0 & 0 & 0 & 2 & 2 & 2 & 2 & 0 & 2 & 1 & 0 & 1 & 0 & 0 & 0 & 1 & 0 & 0 & 2 & 0 & 1 & 2 & 0 & 2 & 1 & 2 & 0\\
        0 & 0 & 0 & 0 & 1 & 0 & 1 & 2 & 2 & 1 & 2 & 0 & 1 & 0 & 2 & 0 & 0 & 0 & 0 & 0 & 0 & 0 & 1 & 0 & 0 & 1 & 0 & 0 & 2 & 2\\
        0 & 0 & 0 & 0 & 0 & 0 & 0 & 1 & 0 & 0 & 1 & 0 & 0 & 2 & 2 & 0 & 0 & 0 & 0 & 1 & 0 & 1 & 0 & 2 & 1 & 0 & 0 & 1 & 2 & 1\\
        0 & 0 & 0 & 0 & 0 & 1 & 1 & 0 & 1 & 0 & 1 & 2 & 1 & 2 & 1 & 0 & 0 & 0 & 0 & 0 & 0 & 0 & 0 & 0 & 1 & 2 & 2 & 1 & 0 & 2\\
        0 & 0 & 0 & 0 & 0 & 0 & 0 & 0 & 0 & 1 & 2 & 2 & 1 & 0 & 2 & 0 & 0 & 0 & 0 & 0 & 1 & 1 & 0 & 1 & 1 & 0 & 1 & 2 & 2 & 0\\
      \end{array}
    \right]
  \end{equation}
  \begin{equation}\label{eq:bigStabilizerBinary}
    \left[
      \begin{array}{*{21}{@{\hspace{.4em}}c}|*{21}{@{\hspace{.4em}}c}}
1 & 0 & 0 & 0 & 0 & 0 & 0 & 0 & 0 & 0 & 0 & 1 & 0 & 1 & 1 & 0 & 0 & 1 & 0 & 0 & 0 & 0 & 0 & 0 & 0 & 0 & 0 & 0 & 0 & 1 & 0 & 1 & 1 & 1 & 1 & 1 & 0 & 1 & 0 & 0 & 0 & 0\\
0 & 0 & 0 & 0 & 0 & 0 & 0 & 0 & 1 & 0 & 1 & 0 & 1 & 1 & 0 & 0 & 0 & 0 & 1 & 0 & 1 & 1 & 0 & 0 & 0 & 0 & 0 & 0 & 0 & 1 & 0 & 1 & 0 & 0 & 1 & 1 & 1 & 0 & 1 & 0 & 0 & 1\\
0 & 1 & 0 & 0 & 0 & 0 & 0 & 0 & 0 & 0 & 0 & 0 & 1 & 0 & 1 & 1 & 0 & 0 & 1 & 0 & 0 & 0 & 0 & 0 & 0 & 0 & 0 & 0 & 0 & 0 & 1 & 0 & 1 & 1 & 1 & 1 & 1 & 0 & 1 & 0 & 0 & 0\\
0 & 0 & 0 & 0 & 0 & 0 & 0 & 0 & 1 & 1 & 1 & 0 & 1 & 1 & 0 & 0 & 0 & 1 & 1 & 1 & 1 & 0 & 1 & 0 & 0 & 0 & 0 & 0 & 0 & 0 & 1 & 0 & 0 & 1 & 0 & 1 & 0 & 0 & 1 & 1 & 0 & 1\\
0 & 0 & 1 & 0 & 0 & 0 & 0 & 0 & 0 & 0 & 0 & 0 & 0 & 1 & 0 & 1 & 1 & 0 & 0 & 1 & 0 & 0 & 0 & 0 & 0 & 0 & 0 & 0 & 0 & 0 & 0 & 1 & 0 & 1 & 1 & 1 & 1 & 1 & 0 & 1 & 0 & 0\\
0 & 0 & 0 & 0 & 0 & 0 & 0 & 0 & 1 & 1 & 0 & 0 & 1 & 1 & 0 & 0 & 0 & 1 & 0 & 1 & 0 & 0 & 0 & 1 & 0 & 0 & 0 & 0 & 0 & 0 & 0 & 1 & 1 & 1 & 1 & 0 & 0 & 1 & 1 & 1 & 1 & 1\\
0 & 0 & 0 & 1 & 0 & 0 & 0 & 0 & 0 & 0 & 0 & 0 & 0 & 0 & 1 & 0 & 1 & 1 & 0 & 0 & 1 & 0 & 0 & 0 & 0 & 0 & 0 & 0 & 0 & 0 & 0 & 0 & 1 & 0 & 1 & 1 & 1 & 1 & 1 & 0 & 1 & 0\\
0 & 0 & 0 & 0 & 0 & 0 & 0 & 0 & 1 & 1 & 0 & 0 & 1 & 0 & 1 & 0 & 0 & 0 & 0 & 0 & 0 & 0 & 0 & 0 & 1 & 0 & 0 & 0 & 0 & 1 & 0 & 1 & 1 & 1 & 0 & 0 & 1 & 0 & 0 & 1 & 1 & 0\\
0 & 0 & 0 & 0 & 1 & 0 & 0 & 0 & 0 & 0 & 0 & 1 & 0 & 1 & 1 & 1 & 0 & 0 & 1 & 0 & 0 & 0 & 0 & 0 & 0 & 0 & 0 & 0 & 0 & 1 & 0 & 1 & 1 & 0 & 1 & 0 & 1 & 0 & 1 & 1 & 0 & 1\\
0 & 0 & 0 & 0 & 0 & 0 & 0 & 0 & 0 & 1 & 1 & 0 & 0 & 1 & 0 & 1 & 0 & 0 & 0 & 0 & 0 & 0 & 0 & 0 & 0 & 1 & 0 & 0 & 0 & 0 & 1 & 0 & 1 & 1 & 1 & 0 & 0 & 1 & 0 & 0 & 1 & 1\\
0 & 0 & 0 & 0 & 0 & 1 & 0 & 0 & 1 & 0 & 1 & 0 & 0 & 1 & 1 & 1 & 1 & 0 & 1 & 1 & 1 & 0 & 0 & 0 & 0 & 0 & 0 & 0 & 0 & 1 & 1 & 1 & 1 & 1 & 1 & 0 & 1 & 1 & 1 & 1 & 1 & 1\\
0 & 0 & 0 & 0 & 0 & 0 & 0 & 0 & 1 & 0 & 0 & 1 & 1 & 1 & 1 & 0 & 1 & 0 & 1 & 0 & 1 & 0 & 0 & 0 & 0 & 0 & 1 & 0 & 0 & 1 & 0 & 0 & 0 & 1 & 0 & 0 & 1 & 0 & 0 & 0 & 0 & 0\\
0 & 0 & 0 & 0 & 0 & 0 & 1 & 0 & 1 & 1 & 1 & 0 & 1 & 0 & 0 & 1 & 1 & 0 & 1 & 1 & 0 & 0 & 0 & 0 & 0 & 0 & 0 & 0 & 0 & 0 & 1 & 1 & 0 & 0 & 1 & 1 & 1 & 0 & 0 & 1 & 1 & 0\\
0 & 0 & 0 & 0 & 0 & 0 & 0 & 0 & 0 & 1 & 0 & 1 & 1 & 0 & 0 & 1 & 0 & 0 & 0 & 1 & 0 & 0 & 0 & 0 & 0 & 0 & 0 & 1 & 0 & 1 & 1 & 1 & 1 & 1 & 0 & 1 & 0 & 0 & 0 & 0 & 0 & 0\\
0 & 0 & 0 & 0 & 0 & 0 & 0 & 1 & 0 & 1 & 1 & 1 & 0 & 1 & 0 & 0 & 1 & 1 & 0 & 1 & 1 & 0 & 0 & 0 & 0 & 0 & 0 & 0 & 0 & 0 & 0 & 1 & 1 & 0 & 0 & 1 & 1 & 1 & 0 & 0 & 1 & 1\\
0 & 0 & 0 & 0 & 0 & 0 & 0 & 0 & 0 & 0 & 1 & 0 & 1 & 1 & 0 & 0 & 1 & 0 & 0 & 0 & 1 & 0 & 0 & 0 & 0 & 0 & 0 & 0 & 1 & 0 & 1 & 1 & 1 & 1 & 1 & 0 & 1 & 0 & 0 & 0 & 0 & 0\\
      \end{array}
    \right]
  \end{equation}
  \caption{Stabilizer matrices for the codes $[\![15,3,5]\!]_3$
    \eqref{eq:bigStabilizer} and $[\![21,5,6]\!]_2$
    \eqref{eq:bigStabilizerBinary}.}
\end{figure*}

\section{New Stabilizer Codes}\label{Sec:New_codes}
We have started a search for quantum stabilizer codes that improve the
lower bounds given in \cite{Grassl:codetables}.  For this, we first
searched the tables in \cite{Grassl:codetables} for codes $[\![n,k,d]\!]_q$ such that a code
$[\![n-t,k,d-t+1]\!]_q$ for some $t>0$ may exist, but has not
yet been found. The following calculations have been executed using
Magma \cite{Magma}.

\begin{itemize}
  \item We start with the code $[\![51,19,9]\!]_2$ from
    \cite{Grassl:codetables}. The corresponding code $S_2^\perp$ is a
    $[51,35,9]_4$ cyclic $\F_4$-linear code which has $6579$
    words of minimum weight. Since $S_2$ is a $[51,16,22]_4$ code, and $S_2\subseteq S_2^\perp$, this is a pure stabilizer code. For the positions $I=\{25,42\}$, we find
    that the pair $(1,\omega)\in\F_4^2$ does not occur. Hence,
    puncturing at those two position using
    $\symp{\vek{\alpha}}{\vek{\beta}}=\symp{1,1}{1,0}$ yields by Proposition \ref{prop:improved_dist} a code
    $[\![49,19,8]\!]_2$ which improves the current lower bound $d\ge
    7$ on the minimum distance in \cite{Grassl:codetables}.  Note that
    using the method for classical codes from \cite{GrasslWhite2004}
    requires to find a set such that every minimum weight vector has a
    zero at a position in that set. In this case, we only find a
    hitting set of size $3$, resulting in no improvement.
    
  \item For the pure code $[\![85,9,19]\!]_2$ from
    \cite{Grassl:codetables}, the corresponding code $S_2^\perp$ is a
    $[85,47,19]_4$ cyclic $\F_4$-linear code which has $78540$
    words of minimum weight.  The automorphism group of this code
    contains a permutation group of order $340$. We use this group to
    reduce the number of $t$-subsets considered for puncturing. For
    $t=4$, we have to consider only $6025$ out of
    $\binom{85}{4}=2024785$ sets. For the set $I=\{1,16,21,27\}$, we
    find that there is no minimum weight word that is equal to the
    tuple $(1,\omega,1,1)$ at those positions. Hence, from Proposition \ref{prop:improved_dist}, puncturing
    yields a code $[\![81,9,16]\!]_2$ which improves the lower bound.
    Note that directly verifying the minimum distance of this code is
    estimated to take more than five years.

    In this case, we find a hitting set for the complement of the
    support of the minimum weight words of $S_2^\perp$ of size
    $5$. Then, shortening the code $S_2$ at those positions (see Section \ref{sec:shortening}) yields the
    stabilizer of a $[\![80,14,15]\!]_2$ quantum code that improves
    the lower bound.

  \item For the pure code $[\![109,37,16]\!]_2$  from
    \cite[Example 5.3]{Dast2024}, the corresponding code $S_2^\perp$ is a
    $[109,73,16]_{4}$ cyclic $\F_{4}$-linear code which has
    $56898$ words of minimum weight. Computing the minimum weight words took more than $5$ CPU years and was completed in about $87$ hours real time using up to $600$ cores.  The automorphism group of this code contains a permutation group of order $1962$. We use this
    group to reduce the number of $t$-subsets considered for
    puncturing.  For $t=3$, it suffices to consider only $114$ out of the $\binom{109}{3}=209934$ subsets. For the set $I=\{1,17,78\}$, we find that there is no minimum weight word that is equal to the
    tuple $(1,\omega,\omega^2)$ at those positions. Hence, puncturing
    yields a code $[\![106,37,14]\!]_2$ according to Proposition \ref{prop:improved_dist} which improves the lower bound.
    Note that directly verifying the minimum distance of this code is
    estimated to take more than $200$ CPU years.
    
    Moreover, there is a hitting set of size $4$. Shortening the code $S_2$ at those positions (see Section \ref{sec:shortening}) yields the
    stabilizer of a $[\![105,41,13]\!]_2$ quantum code that improves
    the lower bound as well.
    
  \item For the pure code $[\![37,1,15]\!]_5$ from
    \cite{Grassl:codetables}, the corresponding code $S_5^\perp$ is a
    $[37,19,15]_{25}$ cyclic $\F_{25}$-linear code which has
    $127872$ words of minimum weight.  The automorphism group of this
    code contains a permutation group of order $666$. We use this
    group to reduce the number of $t$-subsets considered for
    puncturing.  For $t=2$, it is sufficient to consider $\{1,9\}$ and
    $\{1,10\}$.  We do not obtain an improvement puncturing the code
    at $t=1$ or $t=2$ positions.  Instead, we consider symplectic
    self-dual additive subcodes $(37,5^{37},15)_{25}$. There are six
    such codes, and each of them contains $21312=127872/6$ words of
    minimum weight.

    For the set $I=\{1,9\}$, we have $|\mathcal{M}^*_{I,2}|=448<24^2$
    for all six codes. Hence, by Proposition \ref{prop:improved_dist}, in all cases there is a puncturing that
    yields a code $[\![35,0,14]\!]_5$ which improves the lower bound.

\end{itemize}

\section{Conclusion}\label{sec:concl-open-probl}
In this work, we have shown how puncturing of quantum stabilizer codes
can be understood in the language of symplectic vectors. This allows
puncturing to be carried out directly on the stabilizer matrix, thus
giving an explicit description of the generators of the punctured
stabilizer.
Furthermore, our proposed puncturing procedure generalizes the previously
known puncturing methods, in the sense that the additional degrees of freedom are explicitly addressed. They determine which symplectic
vectors are kept and which are removed during puncturing.

This knowledge on puncturing leads to new ways to search for quantum codes,
whose parameters exceed those of previously known codes. In particular, we
discussed such methods in Sections~\ref{sec:Idea_to_good_codes} and \ref{sec:shortening},
and used them to obtain quantum error-correcting codes with parameters
beating the best known codes listed in~\cite{Grassl:codetables}.

In addition, we performed calculations to quantify how often the minimum
distance after puncturing matches the lower bound guaranteed by the
procedure. While the examples are relatively small, we found in these cases
that successive puncturings often produce better minimum distances than the
worst case.

Finally, we have shown how the generalized puncturing procedure allows
transferring the ideas of the classical Griesmer bound to the quantum
setting when $q=p$.

\section*{Acknowledgments}
This paper was supported by the Poul Due Jensen Grant
``Swift'', by the Villum Investigator Grant ``WATER'' grant number 37793 from
the Velux Foundations, Denmark, and by the Danish National Research Foundation (DNRF), through the Center CLASSIQUE, grant number 187.

Grassl's work was carried out under IRA Programme, project
number FENG.02.01-IP.05-0006/23, financed by the FENG programme 2021--2027,
Priority FENG.02, Measure FENG.02.01, with the support of the FNP.
Part of the calculations were carried out on computers funded under
contract number 2018/MAB/5/AS-1, co-financed by EU within the Smart
Growth Operational Programme 2014--2020.

\bibliographystyle{IEEEtran}
\bibliography{Bib}

\vfill

\end{document}